\documentstyle[prd,aps,floats,epsf,psfrag]{revtex}

\newcommand{\mpl}{m_{\mbox{\tiny pl}}}

\newcommand{\be}{\begin{equation}}
\newcommand{\ee}{\end{equation}}

\begin{document}

\draft

%
%

%
\renewcommand{\topfraction}{0.99}
\renewcommand{\bottomfraction}{0.99}

\twocolumn[\hsize\textwidth\columnwidth\hsize\csname 
@twocolumnfalse\endcsname

\title{Nonsingular Dilaton Cosmology}

\author{R. Brandenberger,$^{1}$ Richard Easther$^{1}$ and 
J. Maia$^{1,2,3}$}
\address{1. Department of Physics,  Brown University, 
Providence, RI 02912, USA.}

\address{2. FINPE, Instituto de F\'\i sica, Universidade de S\~ao Paulo, 
CP 66318 05389-970, S\~ao Paulo, SP Brazil. }

\address{3. Departamento de F\'\i sica Te\'orica e Experimental, \\
Universidade
Federal do Rio Grande do Norte, 59072-970, Natal, RN, Brazil.}

\maketitle
\begin{abstract}

We study spatially homogeneous and isotropic solutions to the
equations of motion derived from dilaton gravity, in the presence of a
special combination of higher derivative terms in the gravitational
action. All solutions are nonsingular. For initial conditions
resembling those in the pre-big-bang scenario, there are solutions
corresponding to a spatially flat, bouncing Universe originating in a
dilaton-dominated contracting phase and emerging as an expanding
Friedmann Universe.

\vspace{10pt}
BROWN-HET-1128
\vspace{10pt}

\end{abstract}

]


\section{Introduction}  

The initial singularity is one of the outstanding problems of current
cosmological models. In standard big bang cosmology, the existence of
the initial singularity is an inevitable consequence of the
Penrose-Hawking theorems.$^{\cite{Penrose1965a,Hawking1967a}}$ While
scalar field-driven inflationary models such as chaotic
inflation$^{\cite{Linde1983b}}$ resolve many of the problems faced by
conventional cosmology, initial singularities are still generic, even
when stochastic effects are included.$^{\cite{BordeET1993a}}$ It is
generally hoped that string theory may lead to a resolution of this
problem. In an attempt to address the potential of string theory to
remove the initial cosmological singularity, Gasperini and Veneziano
initiated a program known as pre-big-bang
cosmology$^{\cite{GasperiniET1992b}}$ based on the low energy
effective action resulting from string theory. At lowest order, this
is the action of dilaton gravity,
\be \label{E1}
S = -\frac{1}{2\kappa^2} \int{ d^4 x \sqrt{-g} \left\{
R - \frac{1}{2} (\nabla \phi)^2 + \cdots \right\}},
\ee
where $\phi$ is the dilaton, $\kappa^2 = 8\pi G = 8\pi \mpl^{-2}$,
with $G$ being the (4 dimensional) gravitational coupling, and $\mpl$
the Planck mass.  The field equations of pre-big-bang cosmology
exhibit a new symmetry, scale factor duality, which (in the Einstein
frame) maps an expanding Friedmann-Robertson-Walker (FRW) cosmology to
a dilaton-dominated contracting inflationary phase. This raises the
hope that it is possible to realize a nonsingular cosmology in which
the Universe starts out in a cold dilaton-dominated contracting phase,
goes through a bounce and then emerges as an expanding FRW
Universe.\footnote{See Ref. \cite{Veneziano1998a} for a recent review
of pre-big-bang cosmology.}

Unfortunately, it has been shown that the two branches of pre-big-bang
cosmology cannot be smoothly connected within the tree-level action.%
$^{\cite{BrusteinET1994b,EastherET1995a,KaloperET1995a,KaloperET1995b}}$
The contracting dilaton-dominated branch has a future singularity,
whereas the expanding branch emerges from a past singularity. One-loop
effects in superstring cosmology can regulate the
singularity$^{\cite{EastherET1996a}}$ and smoothly connect a
contracting phase to an expanding phase, at least in the presence of
spatial curvature. Refs
\cite{AntoniadisET1993a,%
LarsenET1996a,%
Rey1996a,%
GasperiniET1996b,%
GasperiniET1996c,%
GasperiniET1997a,%
Bose1997a,%
BoseET1997a,%
KalyanaRama1997a,%
DabrowskiET1997a,%
LukasET1997b,%
BrusteinET1997a,%
Maggiore1997a,%
Gasperini1998a,%
BrusteinET1997c,%
KantiET1998a}
describe other attempts to regulate the singularities of pre-big-bang
cosmology. 

A natural approach to resolving the singularity problem of general
relativity is to consider an effective theory of gravity which
contains higher order terms, in addition to the Ricci scalar of the
Einstein action. This approach is well motivated, since we expect that
any effective action for classical gravity obtained from string
theory, quantum gravity, or by integrating out matter fields, will
contain higher derivative terms. Lastly, in the case of string theory,
t-duality provides further evidence that physical quantities will
remain finite at all times.$^{\cite{BrandenbergerET1989b}}$ Thus, it
is extremely natural to consider higher derivative effective gravity
theories when studying the properties of space-time at large
curvatures.

The various possible extensions to classical general relativity all
hold out the hope of a fully nonsingular theory of gravity, and
therefore a nonsingular cosmology. However, this promise has yet to be
realized in an effective theory of gravity that is rigorously derived
from a well motivated model of Planck-scale physics. In this paper we
approach the problem from a different perspective, and ask instead
whether it is possible to derive a higher order theory of gravity
which will produce a cosmology that captures the qualitative features
of the nonsingular cosmological evolution envisaged by the
pre-big-bang scenario.

An approach to explicitly constructing an effective gravitational
action which insures that physical invariants are always finite is
given in Refs \cite{MukhanovET1992a,BrandenbergerET1993a}. The
resulting action includes a particular combination of quadratic
invariants of the Riemann tensor added to the usual Einstein-Hilbert
action for gravity.  This term forces all solutions of the equations
of motion to approach de Sitter space-time at high curvature, and
therefore renders them nonsingular. The model thus obtained is a
specific higher-derivative gravity theory.

We should mention that a further difficulty with a perturbative
approach to the cosmological singularity problem is that the
perturbation expansion is expected to break down at energy scales
lower than the scale at which the singularity will be smeared out.
Refs~\cite{MukhanovET1992a,BrandenbergerET1993a} also address the
problem of ensuring that there is maximum allowable curvature in a
given cosmological model, using a similar technique to the one
employed to eliminate singularities. At this point, however, we have
chosen to focus our attention on the singularity problem alone, and
have not imposed a limit on the curvature, beyond that implied by the
removal of singularities which ensures that for any given solution the
curvature will be bounded.

The simplest way of achieving a nonsingular cosmology is to add an
invariant $I_2$ to the action, with the property that $I_2 = 0$ is
true if and only if the spacetime is a de Sitter space. By coupling
$I_2$ into the gravitational action via a Lagrange multiplier field
$\psi$ with a potential chosen to ensure that $I_2 \rightarrow 0$ at
large curvatures we can require that all solutions approach de Sitter
space at large curvature, thereby removing the singularity.  For
homogeneous and isotropic spacetimes, a choice for $I_2$ which
satisfies this condition is
\be
I_2 = \sqrt{4 R_{\mu \nu} R^{\mu \nu} - R^2}
\label{I2}
\ee
The simplest way of writing the resulting action is
\be
S(g_{\mu \nu}, \psi) = \int d^4x \sqrt{-g} (R + \psi I_2 
+ V(\psi))
\ee
where $V(\psi)$ is a function chosen such that the action has
the correct (for $\psi \rightarrow 0$) Einsteinian low curvature
limit, whereas for $|\psi| \rightarrow \infty$ the constraint
equation forces $I_2 \rightarrow 0$.

In this paper, we investigate the consequences of adding the same
higher derivative terms to the action of pre-big-bang cosmology, and
examine whether it can eliminate the singularities and produce a
smooth bounce connecting a contracting phase to an expanding Universe.
Our main result is that it is possible to find a potential $V(\psi)$
for the Lagrange multipliers which ensures that all cosmological
solutions of the extended action for dilaton gravity are
nonsingular. Moreover, we show that there is a class of solutions
corresponding to a contracting Universe smoothly connected to an
expanding FRW phase via a bounce. Furthermore, this happens even in
the absence of spatial curvature. Our model, therefore, constitutes a
successful implementation of the goals of pre-big-bang cosmology.

\section{Action and Equations of Motion}

Our starting point is the action (\ref{E1}) for dilaton gravity
(written in the Einstein frame) to which we add the higher derivative
term given by $I_2$, in analogy to what was done in the absence of the
dilaton in Refs. \cite{MukhanovET1992a,BrandenbergerET1993a}:
\be  \label{E4}
S = \frac{-1}{2\kappa^2} \int{ d^4 x \sqrt{-g} \left\{
R - \frac{1}{2} (\nabla \phi)^2 + c \psi e^{\gamma\phi} I_2 + V(\psi) 
\right\}}.
\ee
For the moment, we allow a general coupling between $I_2$ and the
dilaton.  Minimal coupling corresponds to setting the constant
$\gamma$ equal to zero. The constant $c$ rescales the Lagrange
multiplier field $\psi$, and will be chosen to simplify the equations
of motion.

Restricted to a homogeneous and isotropic metric of the form
\be
ds^2 = dt^2 - a(t)^2 \bigl( {1 \over {1 - kr^2}}dr^2 + r^2 d\Omega^2
\bigr) \, ,
\ee
where $d\Omega^2$ is the metric on $S^2$, the equations of motion
resulting from (\ref{E4}) become
\begin{eqnarray}
&\ddot{\phi} + 3 H \dot{\phi} + \gamma c \psi e^{\gamma\phi}\sqrt{12} 
\left(\frac{k}{a^2} - \dot{H} \right) = 0, & \nonumber \\
&\dot{H} =  \frac{k}{a^2} - \frac{e^{-\gamma\phi}}{c \sqrt{12}} 
\frac{\partial V}{\partial \psi}, & \nonumber \\
&6 \frac{k}{a^2} + 6H^2 - \frac{\dot{\phi}^2}{2} - V(\psi) = &\nonumber\\
&c e^{\gamma\phi}\sqrt{12} \left(3 H^2 \psi -\frac{k}{a^2}\psi +
H(\dot{\psi} + \gamma \dot{\phi}\psi)\right), & \label{E6}
\end{eqnarray}
where dots denote derivatives with respect to time, $t$.

Since our initial goal is to construct a spatially flat, bouncing
Universe, we set the curvature constant $k = 0$. We will, for the
moment, consider minimal coupling of $\phi$ to $I_2$, which fixes
$\gamma = 0$. To eliminate useless constant coefficients in the
equations of motion, it is convenient to choose $c \sqrt{12} = 1$. The
resulting equations of motion become
\begin{eqnarray}
\dot{\psi} \, &=& \, - 3 H \psi \, + 6 H \, - \, {1 \over H}
\bigl( {1 \over 2} \chi^2 \, + \, V(\psi) \bigr), \nonumber \\
\dot{H} \, &=& \, \ -V^{\prime}(\psi) , \nonumber \\
\dot{\chi} \, &=& \, - 3 H \chi,  \label{E7}
\end{eqnarray}
with $\chi = \dot{\phi}$ and a prime ($\prime$) signifying derivatives
with respect to $\psi$.

We next turn to a discussion of the criteria which the potential
$V(\psi)$ must satisfy. At small curvatures, the terms in the action
(\ref{E4}) which depend on $\psi$ must be negligible compared to the
usual terms of dilaton gravity. This is ensured by
demanding
\be  
V(\psi) \, \sim \, \psi^2 \,\,\,\,\,\, |\psi| \rightarrow 0 \label{E8}
\ee
as the region of small $|\psi|$ will correspond to the low curvature
domain.$^{\cite{BrandenbergerET1993a}}$ In order to implement the
limiting curvature hypothesis, the invariant $I_2$ must tend to zero,
and thus the metric $g_{\mu \nu}$ will tend to a de Sitter metric at
large curvatures, i.e. for $|\psi| \rightarrow \infty$.  From the
variational equation with respect to $\psi$, it is obvious that this
requires
\be
V(\psi) \, \rightarrow \, {\rm const} \,\,\,\,\,\, |\psi| \rightarrow
\infty \, . \label{E9}
\ee

Conditions (\ref{E8}) and (\ref{E9}) are the same as those employed in
Refs \cite{MukhanovET1992a,BrandenbergerET1993a}, but they do not
fully constrain the potential.  In order to obtain a bouncing solution
in the absence of spatial curvature it is necessary to add a third
criterion: the equations must allow a configuration with $H = 0$ and
$\psi \neq 0$. From the equation of motion for $\psi$ in (\ref{E7}) it
follows that $V(\psi)$ must become negative, assuming that it is
positive for small $|\psi|$. Let $\psi_b$ denote the nontrivial zero
of $V(\psi)$: \be V(\psi_b) \, = \, 0 \, . \label{E10}
\ee In the absence of the dilaton, $\psi_b$ will correspond to the
value of $\psi$ at the bounce. In the presence of $\phi$, the value of
$|\psi|$ at the bounce will depend on $\chi$ and will be larger than
$|\psi_b|$.

A simple potential which satisfies the conditions (\ref{E8}), (\ref{E9})
and (\ref{E10}) is
\be
V(\psi) = \frac{\psi^2 - \frac{1}{16}\psi^4}{1+\frac{1}{32} \psi^4}.
\ee
Note that the potential used in Refs.
\cite{MukhanovET1992a,BrandenbergerET1993a} (which is slightly
simpler) does not satisfy condition (\ref{E10}).

\section{Phase Diagram of Solutions in the Absence of the Dilaton}

The conditions, (\ref{E8}) - (\ref{E10}), on the potential, $V(\psi)$,
discussed in the previous section are necessary but not sufficient to
obtain a nonsingular cosmology. These conditions ensure that all
solutions which approach large values of $|\psi|$ are nonsingular, but
the possibility of geodesic incompleteness for solutions which always
remain within the small $|\psi|$ region remains to be studied. In this
section we determine the phase plane $(\psi, H)$ of our model in the
absence of the dilaton. We use analytical and numerical methods to
study the trajectories of solutions of (\ref{E7}) in the phase plane
and thus explicitly demonstrate the absence of singularities. Our
model therefore yields a further example of a higher derivative
gravity theory without cosmological singularities. In addition, and
unlike the model of Refs. \cite{MukhanovET1992a,BrandenbergerET1993a},
we shall show that our theory admits spatially flat bouncing
solutions.

There are several special points and curves on the phase plane $(\psi, H)$.
First, the point $(\psi, H) = (0, 0)$ corresponds to Minkowski space-time.
The potential $V(\psi)$ vanishes at this point, but it also vanishes at
the points
\be
\psi_b = \pm 4.
\ee
As discussed earlier, the phase plane points $(\psi_b, 0)$ correspond
to bouncing points of cosmological trajectories.

The derivative of $V(\psi)$, and hence ${\dot H}$, vanishes for the
values
\be
\psi_d = \pm 2 \, .
\ee
The phase plane lines $(\psi_d, H)$ are therefore lines along which
${\dot H} = 0$.

To demonstrate that the point $(\psi, H) = (4, 0)$ is in fact a bounce,
we expand the $\psi$ equation of motion near $H = 0$, which yields
\be
H {\dot \psi} \, = \, - V \, . \label{E14}
\ee
Thus, as we cross the $H = 0$ axis, the sign of ${\dot \psi}$ changes.
Contracting solutions with $2 < \psi < 4$ have ${\dot \psi} > 0$ and
approach the point $(4, 0)$ in finite time since ${\dot H}$ is
positive and does not tend to zero, and emerge for $H > 0$ as
expanding trajectories with decreasing curvature (since ${\dot \psi} <
0$). The trajectories in the phase plane are symmetric about the $H =
0$ axis, except that the time arrows are reversed.

Next, we expand the equations near the origin of the phase plane and
obtain
\be
{{d \psi} \over {d H}} \, \simeq \, {1 \over {2 H}}
\bigl( \psi \, - \, {{6 H^2} \over {\psi}} \bigr) \, ,\label{E15}
\ee
from which we see immediately the existence of critical lines located at
\be
\psi_c(H) \, = \, \pm \, \sqrt{6} H \, .
\ee
Focusing on the contracting solutions, the equation of motion for
$\dot{H}$ reduces to 
\be
{\dot H} \, \simeq \, - 2 \psi. \label{E17}
\ee
Consequently, trajectories which lie above the critical line have
${\dot H} < 0$ and (from Eq. (\ref{E15})) ${\dot \psi} > 0$. These
trajectories thus are directed towards the line $\psi = 2$ where
${\dot H}$ changes sign. Provided they do not cross the critical line,
solutions which start out in this region of the phase plane are thus
candidates for spatially flat bouncing Universes.

\begin{figure}[tbp]
\begin{center}
\begin{tabular}{c}
\epsfxsize=8cm 
\psfrag{xl}[bt]{$\psi$}
\psfrag{yl}{$H$}
\epsfbox{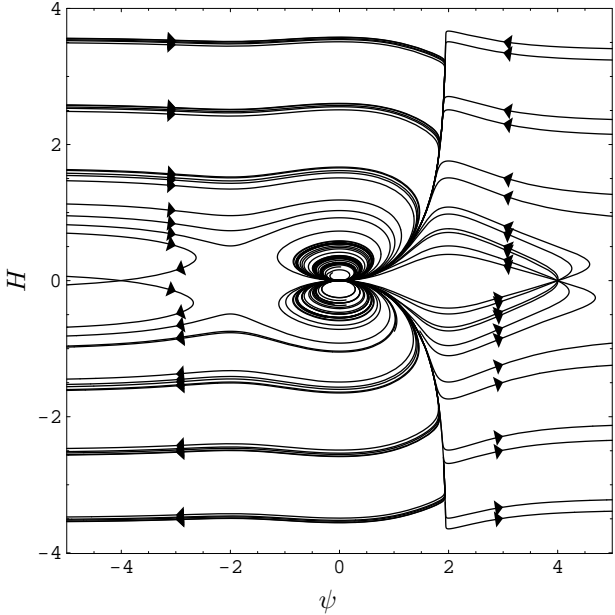}
\end{tabular}
\end{center}
\caption[fig1]{The phase portrait for solutions of the equations of
motion, with the dilaton kinetic energy ($\chi$) set to zero is
shown. The Hubble parameter, $H$, is plotted on the vertical axis,
while $\psi$ is plotted on the horizontal.}
\end{figure}

Solutions below the critical line have ${\dot \psi} < 0$ and will
hence not exhibit a bounce. Note that the critical line is not
itself a trajectory of the dynamics. Indeed, on the critical line
trajectories point in vertical direction, since $d\psi / dH = 0$.

There is a separatrix line between phase plane trajectories which
start near the origin and which reach $\psi = 2$ (and are thus
candidates for a bouncing Universe) and those trajectories which cross
the critical line $\psi_c(H)$ and turn around, i.e. become solutions
with ${\dot \psi} < 0$. To determine the location of the separatrix
line, we solve the equations (\ref{E7}) near the origin of the phase
plane for $|H| \ll \psi$, in which case the variational equation with
respect to $\psi$ in (\ref{E7}) becomes
\be H {\dot \psi} \, \simeq \, - \psi^2 \, , \label{E18}
\ee
which must be solved together with (\ref{E17}). After differentiating
(\ref{E17}) with respect to time and substituting (\ref{E18}) to
eliminate ${\dot \psi}$, we obtain the equation 
\be \label{E19} H
{\ddot H} \, - \, {1 \over 2} {\dot H}^2 \, = \, 0 \, ,
\ee 
which has the solution 
\be 
\label{E20}
H(t) \, = \, -c t^2 \, ,
\ee
where $c$ is a constant which labels the different trajectories. The
second solution of (\ref{E19}) is $H(t) = {\rm const}$ and is thus
uninteresting. From (\ref{E17}) it follows that for the above solution
\be \label{E21}
\psi(t) \, = \, c t \, .
\ee
Equations (\ref{E20}) and (\ref{E21}) give the solutions of the
dynamical equations near the origin of the phase plane, provided that
$|H| \ll \psi$. To get an idea for where the separatrix line lies, we
demand that the trajectory be above the critical line at $\psi =
2$. This condition becomes
\be \label{E22}
|H| \, = \, {1 \over c} \psi^2 \,\,\, {\rm with} \, c \geq 4
\sqrt{2}
\, .
\ee
Setting $c = 4 \sqrt{2}$ in (\ref{E22}) gives a first estimate for the
location of the separatrix line.  Since the set of phase plane
trajectories near the origin is labeled by the parameter $c$, it
follows from (\ref{E22}) that the set of contracting Universes
starting near Minkowski space-time which are candidates for a
spatially flat bouncing Universe has finite measure.

When $\psi = \pm 2$, ${\dot H}$ changes sign. Contracting solutions
which begin near the origin in the phase plane reach a maximal value
of $|H|$ at $\psi = 2$. What happens to these trajectories next
depends on the value of $|H|$ for $\psi = 2$. If
\be \label{E23}
|H(\psi = 2)| \ll 1 \, ,
\ee
then (\ref{E14}) gives a good approximation to the dynamics in the
region $2 < \psi < 4$, and we conclude that the
trajectories bounce. Making use of $|H| = \psi^2 / c$ (see
(\ref{E22})), the condition (\ref{E23}) becomes
\be \label{E24}
c \gg 4 \, ,
\ee
which is consistent with the previous condition (\ref{E22}) for
bouncing trajectories.

Trajectories with $|H(\psi = 2)| \gg 1$ also reach their maximal value
of $|H|$ at $\psi = 2$, but they do not bounce because the $\psi$
equation of motion can now be approximated by
\be \label{E25}
{\dot \psi} \, \simeq \, - 3 H \psi \, ,
\ee
which implies that $\psi$ keeps growing indefinitely. In combination
with the equation of motion for $H$, which yields that
\be \label{E26}
{\dot H} \, \simeq \, 0 \,\,\,\, {\rm for} \, |\psi| \gg 4 \, ,
\ee
we see that the solutions tend to contracting de Sitter
ones. Solutions below the critical line have ${\dot \psi} < 0$ and
will not lead to a bouncing Universe. 

For values of $\psi$ approaching $\psi = 2$, the two critical lines
diverge to $|H| \rightarrow \infty$. This can be seen immediately from
the $\psi$ equation of motion in (\ref{E7}).  For large values of
$|H|$, the right hand side of this equation is dominated by
the first two terms. In the absence of the third term, the solution of
${\dot \psi} = 0$ which determines the critical line would be
$\psi = 2$. The third term, however, leads to a small but
positive correction to the first term (in the bottom right quadrant of
the phase plane which we are considering throughout), thus shifting
the critical line slightly to the left of $\psi = 2$, satisfying
\be \label{E27}
- 3 H \psi \, + \, 6 H \, - \, {1 \over H} V \, = \, 0 \, .
\ee

To prove that all solutions of the equations (\ref{E7}) are
nonsingular, we must show that the solutions asymptotically tend to
either the origin of the phase plane (Minkowski space-time), or to de
Sitter space (the regions $|\psi| \rightarrow \infty$). Solutions
which have $|H| \rightarrow \infty$ for finite $|\psi|$ must be
excluded. The regions along the critical lines are the only
part of the phase plane where there is the danger of singular
trajectories.  However, differentiating the $\psi$ equation of motion
with respect to time, evaluating the result along the critical line
and making use of (\ref{E27}), we find
\be \label{E28}
{\ddot \psi} \, = \, - V^{\prime} {{2 V} \over {H^2}} \, < \, 0 \,  .
\ee
Hence, the critical line is not an attractor, but rather trajectories
peel away from the line and tend to the asymptotic de Sitter region.

We have also solved the equations of motion (\ref{E7})
numerically. The resulting phase diagram is shown in Figure 1. The
absence of singular solutions is manifest. In the asymptotic regions
$|\psi| \gg 4$, all solutions tend to de Sitter space. The critical
lines are seen to repel contracting solutions towards the asymptotic
de Sitter regions. The most interesting class of solutions are those
corresponding to a cosmological bounce. As predicted, they form a set
of finite measure among solutions which start out close to the origin
of the phase plane. There are also solutions which ``oscillate'' about
Minkowski space-time. These solutions are further discussed in the
following section, since we expect that they will be strongly
perturbed by the presence of matter, in particular by the dilaton.

\section{Effects of the Dilaton on the Phase Space Trajectories}

\begin{figure}[htbp]
\begin{center}
\begin{tabular}{c}
\epsfysize=5cm 
\psfrag{xl}{$t$} 
\psfrag{yl}{$H$}
\psfrag{tag}{(a)} 
\epsfbox{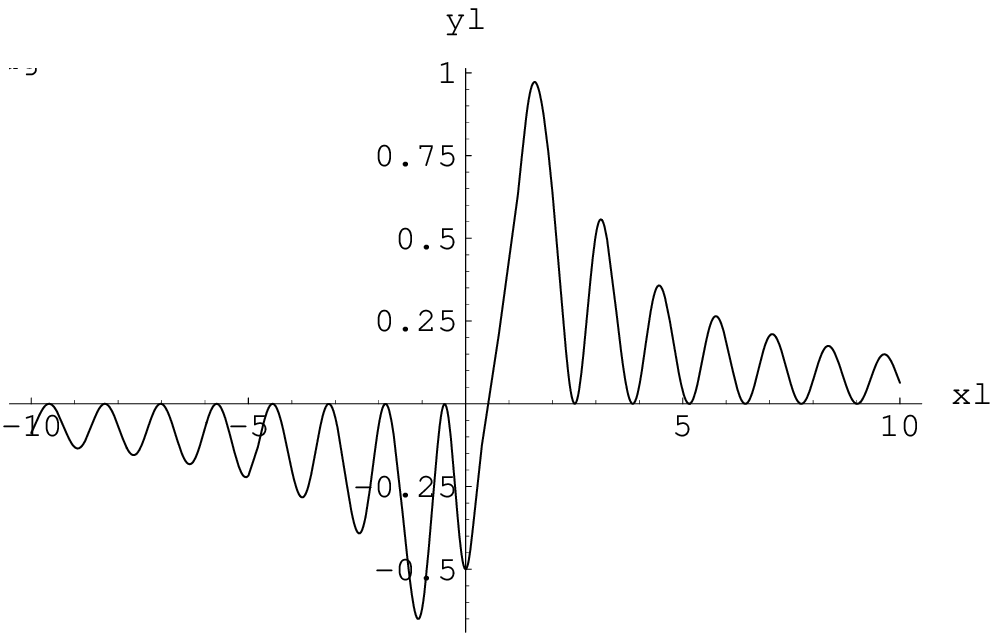} \\
\epsfysize=5cm 
\psfrag{xl}{$t$} 
\psfrag{yl}{$a$}
\psfrag{tag}{(b)} 
\epsfbox{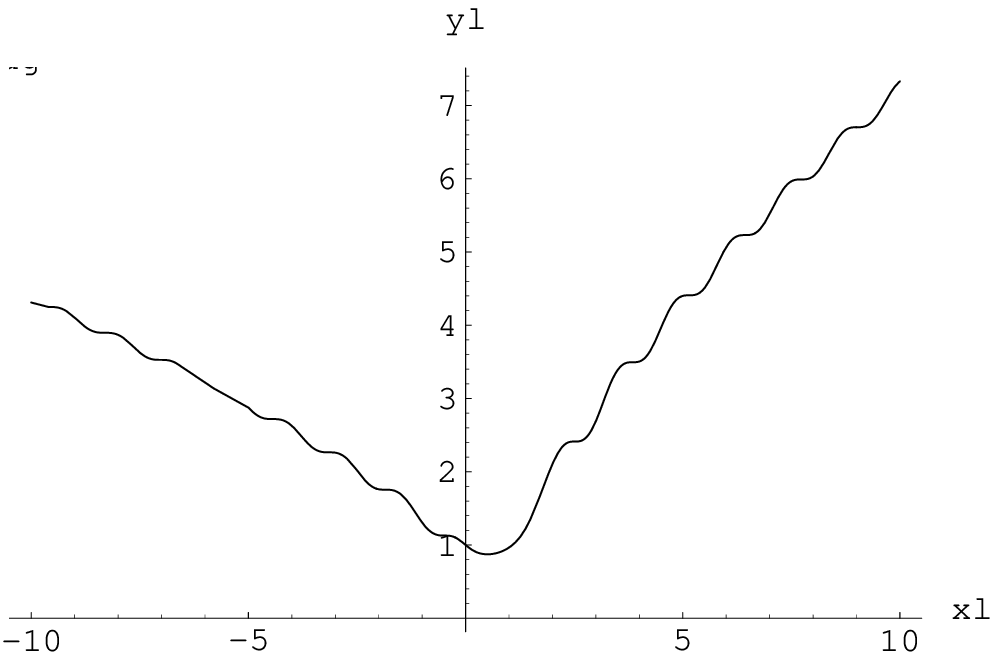} \\
\epsfysize=5cm 
\psfrag{xl}{$t$} 
\psfrag{yl}{$\chi$}
\psfrag{tag}{(c)} 
\epsfbox{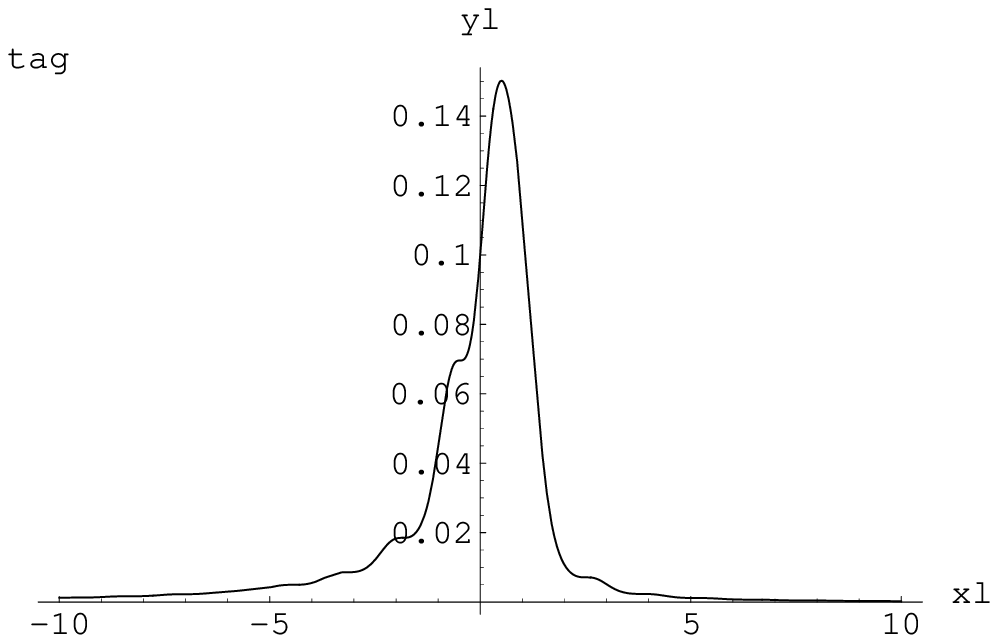} \\
\epsfysize=5cm 
\psfrag{xl}{$t$} 
\psfrag{yl}{$\psi$}
\psfrag{tag}{(d)} 
\epsfbox{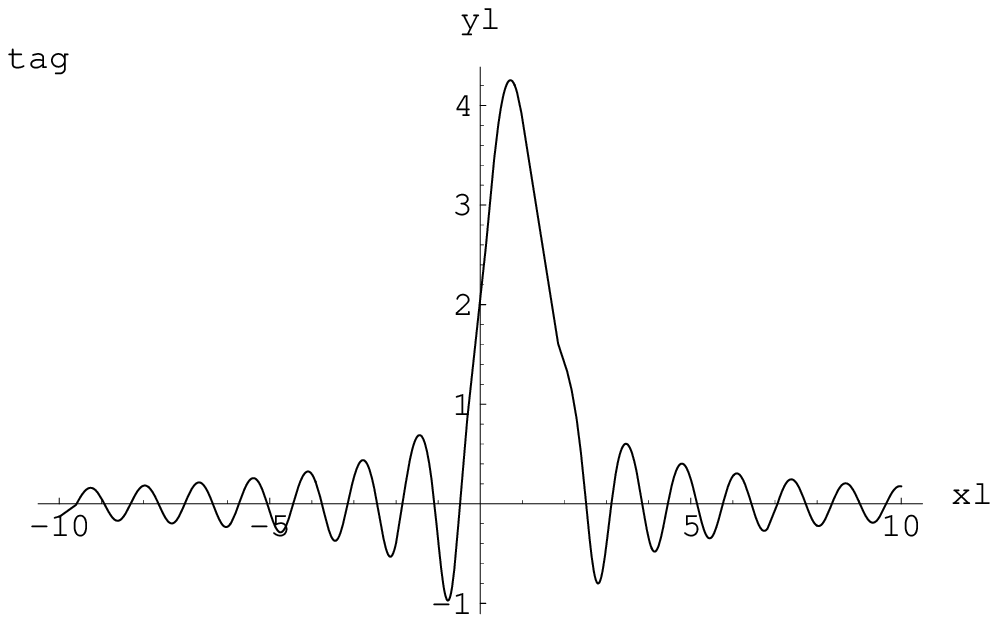} 
\end{tabular}
\end{center}
\caption[fig2]{A specific bouncing solution with a non-trivial
contribution from the dilaton is plotted. Panel a) shows the evolution
of $H$ as a function of time, Panel b) depicts that of $a(t)$, Panel
c) that of $\chi(t)$ and Panel d) that of $\psi(t)$.}
\end{figure}

In the presence of the dilaton, the phase space becomes three
dimensional ($\psi(t) , \,  H(t) , \, \chi(t)$) and therefore more
difficult to discuss analytically. We first note that in the Einstein
frame, the dilaton corresponds to a homogeneous free massless scalar
field with equation of state
\be \label{E29}
p \, = \, \rho \, ,
\ee
where $p$ and $\rho$ denote pressure and energy density, respectively.
This implies
\be \label{E30}
\rho(t) \, \propto \, a(t)^{-6}
\ee
or $N = 6$ in the notation of \cite{BrandenbergerET1993a}. Thus, it is
already clear from the analysis of \cite{BrandenbergerET1993a} that
the dilaton will not introduce any singularities into the system. In
fact, from the equation of motion for $H$ (see (\ref{E7})) it follows
that, in the region of large $|H|$, the presence of $\chi$ will not
change the phase space trajectories projected onto the ($\psi, H$)
plane (a property called ``asymptotic freedom'' in
\cite{BrandenbergerET1993a}). However, the presence of $\chi$ will
greatly accelerate the time evolution of $\psi$ on the given ($\psi,
H$) trajectory. This is easy to see for the contracting de Sitter
solutions, since in this case the $\chi$ equation of motion (see
(\ref{E7}) leads to exponential growth of $\chi$ which, inserted into
the $\psi$ equation of motion, demonstrates that at large values of
$|\psi|$, the $\chi^2$ term dominates the evolution of $\psi$.

We next study the effect of $\chi$ on the trajectories in the ($\psi,
H$) phase plane. The role of the lines $\psi = \pm 2$ remains
unchanged: they correspond to maxima of $|H|$ for any given
trajectory. However, the condition for the bounce changes. Instead of
$V(\psi_b) = 0$, it now follows from (\ref{E7}) that the condition
becomes
\be \label{E31}
{1 \over 2} \chi^2 \, + \, V(\psi_b) \, = \, 0 \, .
\ee
Hence, $|\psi_b|$ is shifted to a larger (and dilaton-dependent)
value. Note that for very large values of $|\chi|$ for which the
above equation has no solutions there will be no bounce. 

The presence of $\chi$ also changes the critical lines. For a fixed
value of $\psi$, the condition ${\dot \psi} = 0$ which determines the
critical line occurs at a larger value of $|H|$ than it does in the
absence of $\chi$, as can easily be seen from the $\psi$ equation of
motion (see (\ref{E7}).  In addition, for fixed initial ($\psi, H$)
the value of ${\dot \psi}$ for contracting solutions is larger with
$\chi \neq 0$ than with $\chi = 0$, so we conclude that adding a small
value of $\chi$ increases the range of initial conditions in the
($\psi, H$) plane which lead to a bounce. Furthermore the larger the
value of $\chi$, the larger the effect is (as long as (\ref{E31})
still has a solution).

We will now argue that collapsing solutions in the presence of a
small $|\chi|$ quite generically lead to a spatially flat bouncing
Universe. Consider initial conditions with small but
positive $\chi$ which in the ($\psi, H$) plane lie below the
separatrix line and which initially ``oscillate'' about Minkowski
space-time. Note that since 
\be \label{E32} 
{\ddot H} \, = \, - V^{\prime} {\dot \psi} \, , 
\ee 
and since ${\dot \psi} > 0$ for these trajectories, they do not bounce
upon reaching $H = 0$ but start another cycle with $H < 0$. Since $H
\leq 0$ at all times, $\chi(t)$ is increasing. Eventually, $\chi(t)$
reaches a sufficiently large value such that the trajectory crosses
the ``separatrix'' in the ($\psi, H$) plane and evolves past $\psi =
2$ to a successful bounce.  An example of such a trajectory is shown
in Figure 2.

It is now obviously possible to construct initial conditions in which
the Universe is initially dominated by the dilaton and contracting
towards a bounce. For example, we can take the initial conditions for
($\psi, H, \chi$) to be those of the solution in Figure 2 at the end
of the cycle preceding the bounce. We have thus shown that our model
provides a successful implementation of the evolution postulated in
pre-big-bang cosmology. Note that after the bounce, the dilaton
rapidly becomes irrelevant to the evolution in the expanding phase.
Note, in particular, that the dilaton tends to a constant. This
feature is different from what happens in other attempts to achieve a
branch change where the dilaton continues to grow in the expanding
phase.

\section{Conclusions}

We have shown that the ``limiting curvature construction'' of
Refs. \cite{MukhanovET1992a,BrandenbergerET1993a} can be applied to
dilaton cosmology to yield spatially flat bouncing cosmological
solutions. In particular, this enables us to implement a model which
successfully implement the qualitative evolution associated with a
branch change in pre-big-bang cosmology.

Specifically, we have investigated spatially homogeneous, isotropic,
and flat solutions of the equations of motion for dilaton gravity
coupled to, $I_2$, a special combination of invariants quadratic in
the Riemann curvature. The coupling is provided by a Lagrange
multiplier field $\psi$. We have determined the conditions which must
be imposed on the potential $V(\psi)$ required in order to guarantee
that all solutions are singularity-free, and to permit bouncing
scenarios.

We then studied the phase space of trajectories and demonstrated, both
analytically and numerically, that all the solutions to the equations
of motion which follow from our action are nonsingular. We identified
a large class of contracting trajectories which start out near the
origin in phase space and develop a cosmological bounce. The addition
of the dilaton increases the fraction of phase space for which
solutions bounce. These bouncing cosmologies display the qualitative
behavior associated with the pre-big bang branch change.

An interesting result of our work is the conclusion that the dilaton is
not necessary to obtain a spatially flat bouncing cosmology. Also, in
many cases a dilaton-dominated bounce is preceded by a period when the
dilaton has a negligible effect on the dynamics of
$\psi(t)$ and $H(t)$.

Our work does not address the issue fixing the value of the dilaton at
late times.  Nor does it address recent objections
${\cite{KaloperET1998a,Coule1997a,TurnerET1997a}}$ to the pre-big-bang
scenario. In work in progress, we are investigating the possibility
that our implementation of the branch change will alleviate some of
these difficulties, in particular the flatness problem and the problem
of how to obtain a large Universe. The relevant feature of our model
which we expect to play an important role in addressing these problems
is that the dynamics of our bounce are determined by the higher
derivative gravity terms which lead to de Sitter phases for many
trajectories. Thus, on balance, our model looks quite promising.

The obvious drawback of our method is that the extra terms in the
action are put in by hand rather than being derived from fundamental
physics.  There are, however, several justifications for following our
approach.  First, it is generally expected that physical invariants
must be limited in string theory. This should, at high curvatures, be
reflected in the effective action for gravity, and our action is an
easy way to realize limiting quantities. Secondly, higher derivative
terms will inevitably arise in effective theories of gravity, such as
those derived from string theory, from other approaches to quantum
gravity, or from quantizing matter fields in curved space-time.  Our
action, even though not directly derived from string theory,
demonstrates that a specific set of higher curvature corrections can
both ensure that all relevant physical invariants remain finite, and
produce a bouncing nonsingular cosmology.

\section*{Acknowledgments}
RB and RE are supported by the DOE contract DE-FG0291ER40688 (Task A).
JM was supported by CAPES, and thanks the High Energy Theory group at
Brown University for its hospitality.  Computational work in support of
this research was performed at the Theoretical Physics Computing
Facility at Brown University. We thank M. Gasperini and G. Veneziano
for their comments on a draft of this paper.


\end{document}